\documentclass[prl,twocolumn]{revtex4}
\usepackage[dvips]{graphicx}
\usepackage{epsfig}
\usepackage{amsmath,amsfonts,amssymb,bm}

\begin{document}
\title{Universal dielectric loss in glass from simultaneous bias and microwave fields}
\author{Alexander L. Burin} 
\affiliation{Department of Chemistry, Tulane University, New
Orleans, LA 70118, USA}
\author{Moe S. Khalil and Kevin D. Osborn}
\affiliation{Laboratory for Physical Sciences (LPS), 8050 Greenmead Drive, 
College Park, MD 20740}
\date{\today}
\begin{abstract}
We derive the ac dielectric loss in glasses due to resonant processes created by two-level systems and a swept electric field bias.
It is shown that at sufficiently large ac fields and bias sweep rates the nonequilibrium loss tangent created by the two fields approaches a universal maximum determined by the bare linear dielectric permittivity. In addition this nonequilibrium loss tangent is derived for a range of bias sweep rates and ac amplitudes and show that the loss tangent creates a predicted loss function that can be understood in a Landau-Zener theory and which can be used to extract the TLS density, dipole moment, and relaxation rate.
\end{abstract}

\maketitle

\begin{figure}[h!]
\centering
\includegraphics[width=5cm]{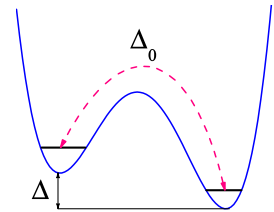}
\caption{ The potential to a tunneling two-level system in an amorphous solid. $\Delta$ is the energy difference between the left and right well states when isolated, which are coupled with tunneling amplitude $\Delta_0$.}
\label{fig:Fig1TLS}
\end{figure}

Low temperature two level systems (TLS) in amorphous solids have attracted growing attention recently due to their performance limiting effects in superconducting qubits and resonators for quantum computing \cite{Martinec,Ustinov,KO1} and kinetic inductance photon detectors for astronomy \cite{Gao}.
TLS, commonly represented by atoms or groups of atoms tunneling between two configuration states (see Fig. \ref{fig:Fig1TLS}, \cite{AHVP}), are found to decrease the coherence of qubits due to microwave absorption. Amorphous dielectrics with TLS are commonly found in Josephson junction barriers and wiring crossovers \cite{Ustinov,KO1}, but TLS are even found on the surfaces of the resonators with coplanar superconducting electrodes on crystalline substrates \cite{Gao}.



Despite theoretical and experimental studies of amorphous solids over recent decades \cite{Hunklinger,Phillips,ab1}, the identification of the tunneling systems and an understanding of their non-equilibrium phenomena have limited our ability to predict their influence on devices. In previous nonequilibrium studies the application of a dc bias voltage step with a small ac field frequency $\omega \ll \frac{k_{B}T}{\hbar}$, results in a time dependent permittivity which indicates complex behavior of the TLS \cite{O1}. Initially after the electric field step, the permittivity increases quickly followed by a slow, logarithmic, return to the equilibrium value. The observed behavior is interpreted as a consequence of the bias field induced change in the TLS density of states caused by long-range TLS interactions \cite{ab1,abJLTP}. A numerical treatment of this low frequency phenomena compares the result to interacting and non-interacting TLS theory  \cite{Clare2}.



A recent experiment measures the high frequency ($\hbar\omega \gg k_{B}T$) ac loss of dielectric films with a simultaneous parallel electric field bias. The study adjusts the sweep rate of the bias and the ac field amplitude such that they are able to probe various nonequilibiurm regimes of the films, and this provides an opportunity to understand nonequilibrium glass properties in the high frequency regime. The bias sweeps are created by applying different dc voltage steps to the film through low-pass filters with time constants which are large compared to the resonator response time and the relaxation time of the TLSs.




In high frequency steady state measurements of amorphous dielectrics the resonant loss caused by TLS lowers (saturates) at large ac fields because the TLS Rabi frequency is larger than the TLS relaxation rate  \cite{Martinec,KO1,Hunklinger2}. However, in the nonequilibrium study with a large ac field the addition of a sufficiently large bias field sweep rate causes the loss tangent to increase back to the value found for the small ac field steady state \cite{K2}.

 



In this paper we will show that the experimentally-observed maximum in nonequilibrium loss tangent should universally reach the value found for small ac fields in the steady state since it can be derived from the conventional tunneling model of amorphous solids. In addition we will derive the loss for a range of ac fields and bias rates and show that it follows a theory which depends on Landau-Zener transitions. The derivation of the nonequilibrium loss allows one to predict the TLS density, dipole moment and relaxation rate for a given experimental result.


Each TLS can be characterized by its asymmetry, $\Delta$ and tunneling amplitude $\Delta_{0}$ \cite{AHVP}, which determine  its energy, $E=\sqrt{\Delta^{2}+\Delta_{0}^2}$. The TLS distribution is given by the universal law, $P(\Delta, \Delta_{0})=\frac{P_{0}}{\Delta_{0}}$, reflecting the exponential sensitivity of the tunneling amplitude to the two-well configuration.  TLS interaction with an external electric field, $\mathbf{F}$, is determined by its dipole moment $\mathbf{p}$ and contributes to the asymmetry energy $\Delta(\mathbf{F})=\Delta(0)- 2\mathbf{p}\mathbf{F}$.  
Following \cite{Martinec} we assume the TLS dipole moment to be a randomly oriented vector with a fixed size $p\sim 1-10$D, 
which is assumed in the majority of theoretical models (e. g. \cite{Hunklinger2}). 
The use of different models, e.g. a Gaussian distribution of dipoles will not change the main result, i.e. the universal value of the loss tangent obtained at large ac fields and bias rates. The loss tangent obtained here as a function of the bias rate and ac field amplitude depends on the dipole size, such that the loss tangent obtained in this work can be generalized for a distribution of dipoles. The interaction of TLS with the environment, i.e. phonons and other TLS,  are characterized by its relaxation time $T_1$ and coherence time $T_2$ \cite{Hunklinger}.


In the absence of the external dc bias field the interaction of TLS with the ac field $\mathbf{F}_{ac}$ results in a field energy absorption which can be described by the loss tangent defined as $\tan(\delta)=\frac{\epsilon''}{\epsilon'}$, where $\epsilon=\epsilon'+i\epsilon''$ is the complex permittivity. We consider the regime of a very high ac field frequency, $\omega \gg \frac{k_{B}T}{\hbar} \gg \frac{1}{T_{1}}, \frac{1}{T_{2}}$, respectively, which takes place for a typical experimental frequency $\nu \sim 5$GHz ($\frac{h\nu}{k_{B}}\sim 200$mK) and a very low temperature $T\leq 30$mK \cite{Martinec,K2}, considered in this letter. In this regime relaxation and decoherence for resonant TLS are determined by spontaneous decays of excited states, so that  $T_{2} \approx 2T_{1}$ \cite{QuantOptics}.

The spontaneous emission via photons can modify the ordinary decay via phonons in a given experiment, and both will contribute to relaxation. However, the coupling of the thermal bath is assumed to be sufficiently high such that the TLS far from resonance are initially in the ground state. As described below, this results in a TLS that will irreversibly (spontaneously) emit if coherently excited after crossing through resonance with the ac field, such that the coherent deexcitation processes at the crossing can be ignored. 


The contribution of a given TLS to the dielectric losses is determined by its interaction with the external microwave field, $F_{ac}\cos(\omega t)$, and environment, which are defined by the TLS Rabi frequency $\Omega_{R}\sim \frac{pF_{ac}}{\hbar}$ and relaxation time $T_{1}$. 
In the resonant regime, with $\Omega_{R} \ll \omega$ and negligible ac Stark effects \cite{Datta}, one can express the TLS interaction Hamilonian $\widehat{h}$ using the rotating frame approximation as
\begin{eqnarray}
\widehat{h}=-\hbar\Omega_{R}\widehat{S}^{x}-(E-\hbar\omega)\widehat{S}^{z}, 
~ \Omega_{R}=\Omega_{R0}\cos(\theta)\frac{\Delta_{0}}{E},
\nonumber\\
~ \hbar\Omega_{R0}=pF_{ac},
~ T_{1}=T_{1,min} \left(\frac{E}{\Delta_{0}}\right)^{2}. 
\label{eq:indiv_polaris}
\end{eqnarray}
Here the minimum relaxation time $T_{1,min}$ and maximum Rabi frequency $\Omega_{R0}$ are defined at $\Delta_0=E$ and the latter also uses a zero for the angle $\theta$ between the dipole $\bold{p}$ and the ac field $\bold{F}_{ac}$.  

Solving the Bloch equations for the TLS density matrix \cite{Clare2} and 
averaging the result over the TLS distribution $P(\Delta, \Delta_{0}) = \frac{P_{0}}{\Delta_{0}}$  results in the following expression for the loss tangent \cite{Martinec,Hunklinger,Hunklinger2}
\begin{equation}
\tan(\delta)\approx\frac{4\pi^2 P_{0}p^2\tanh\left(\frac{\hbar\omega}{2k_{B}T}\right)}{3\epsilon\sqrt{1+\Omega_{R1}^2 T_{1,min}T_{2,min}}}, 
\label{eq:losstan_eq}
\end{equation}
where where $\epsilon$ is the relative permittivity (permittivity in cgs units, in our case of interest $T_{2,min}=2T_{1,min}$). Here the effective Rabi frequency after averaging over the angle $\theta$ is $\Omega_{R1}\approx \frac{8}{3\pi}\Omega_{R0}$. The high field saturation of the loss tangent is caused by a TLS Rabi oscillations which are faster than the relaxation to the environment. The temperature dependence of the steady state loss reflects the thermal dependence of the TLS population difference.




For a swept bias field parallel to the ac field the energy of the TLS depends on time as $E(t)=\sqrt{(\Delta-2\mathbf{p}\mathbf{F}_{bias}(t))^2+\Delta_{0}^2}$. A TLS contributes to the microwave absorption near the resonance $E\approx \hbar\omega$. Then one can approximately represent the energy of a TLS in the resonant form 
\begin{eqnarray}
E(t)=\hbar\omega+\hbar v(t-t_{0}),
\nonumber\\ 
v = v_{0}\sqrt{1-\left(\frac{\Delta_{0}}{\hbar\omega}\right)^2}\cos(\theta),
~ \hbar v_{0} = 2p\frac{dF_{bias}}{dt}
\label{eq:E_of_t}
\end{eqnarray}
where $t_{0}$ defines the time when the exact resonance takes place. Two resonances are generally possible for each TLS $\Delta-2\mathbf{p}\mathbf{F}_{bias}(t)=\pm\sqrt{(\hbar\omega)^2-\Delta_{0}^2}$. We assume that the time between two resonant passages $\frac{\omega_{0}}{v_{0}}$ is longer than the TLS relaxation time $T_{1,min}\ll \omega$ so that we can treat them independently.

\begin{figure}[h!]
\centering
\includegraphics[width=5.5cm]{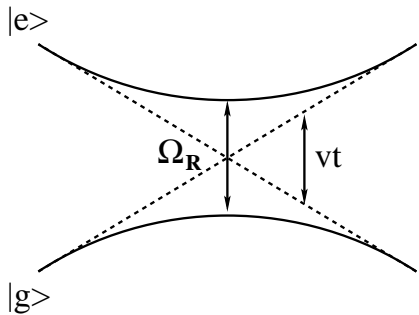}
\caption{TLS energy spectrum as a function of time, induced by the bias field application. The ground and excited states are coupled by one photon transitions, described within the rotating wave approximation.}
\label{fig:Fig2LZ}
\end{figure}

We begin the consideration of the non-equilibrium loss tangent with the oversimplified case when the relaxation and decoherence rates are small during resonance passage and can be neglected. Then the loss will take place after the resonance crossing events induced by the ac and bias fields (see Fig. \ref{fig:Fig2LZ}). In this regime TLS can be described by the wave function amplitudes in the ground and excited states $(c_1,c_2)$, respectively. The modified wave function $(a_{1}, a_{2})=(c_{1}e^{i\omega t/2}, c_{2}e^{-i\omega t/2})$ taken within the rotating frame approximation Eq. (\ref{eq:indiv_polaris}) 
satisfies equations, 
\begin{eqnarray}
\frac{da_{1}}{dt}=i\frac{v(t-t_{0})}{2}a_{1}-i\frac{\Omega_{R}}{2}a_{2},
\nonumber\\ 
\frac{da_{2}}{dt}=-i\frac{v(t-t_{0})}{2}a_{2}-i\frac{\Omega_{R}}{2}a_{1}, 
\label{eq:Schr1}
\end{eqnarray}
which are equivalent to Landau-Zener transition dynamics of a two-level quantum system (see Fig. \ref{fig:Fig2LZ}, \cite{LZ}).
 If at $t=-\infty$ only the ground state is populated $\mid a_{1}\mid^2=1$, $\mid a_{2}\mid^2=0$ then after the level crossing, $t=\infty$, one has $\mid a_{1}\mid^2=\exp\left(-\frac{\pi\Omega_{R}^2}{2v}\right)$, $\mid a_{2}\mid^2=1-\exp\left(-\frac{\pi\Omega_{R}^2}{2v}\right)$. 

The `imaginary part of TLS dielectric constant $\epsilon''$ is defined by the reactive component of the TLS dipole moment, $\mathbf{p}_{loss}$, which  can be expressed as \cite{Clare2}
\begin{equation}  
\mathbf{p}_{loss}=-i\frac{\Delta_{0}}{\hbar\omega}\mathbf{p}(a_{1}^{*}(t)a_{2}(t)-a_{2}^{*}(t)a_{1}(t)). 
\label{eq:ImPolarisTLS}
\end{equation}
This expression should be averaged over TLS parameters including their energies, $E=\hbar(\omega+v(t-t_{0}))$, tunneling amplitudes $\Delta_{0}$ and dipoles $\mathbf{p}$. The integration over energy can be performed analytically employing the fact that $(a_{1}^{*}(t)a_{2}(t)-a_{2}^{*}(t)a_{1}(t))=i\frac{d}{\Omega_{R}dt}(\mid a_{1}\mid^2-\mid a_{2}\mid^2)$. Then this integration of Eq. (\ref{eq:ImPolarisTLS}) is equivalent to the integration over time, $dE = vdt$. It results in the Landau-Zener change in population difference, $\frac{2|v|}{F_{ac}}\left[1-\exp\left(-\frac{\pi\Omega_{R}^2}{2v}\right)\right]$.
The average TLS loss tangent can then be written in terms of the previous expression integrated over the distribution in tunneling amplitudes
\begin{equation}
\tan(\delta)= \frac{8\pi P_{0}}{\epsilon F_{ac}^2}\int_{0}^{\hbar\omega} \frac{d\Delta_{0}}{\Delta_{0}}\frac{\left<\hbar^{2}|v|\left(1-e^{-\frac{\pi\Omega_{R}^2}{2|v|}}\right)\right>}{\sqrt{1-\left(\frac{\Delta_{0}}{\hbar\omega}\right)^{2}}}, 
\label{eq:losstan_neq}
\end{equation}
where $<...>$ indicates an average taken over the dipole direction with respect to the field.

In the limit of a large bias sweep  rate, 
$\Omega_{R0}^2 \ll v_{0},$
one can approximate the exponent in Eq. (\ref{eq:losstan_neq}) as $e^{-x}\approx 1-x$. Then all integrals can be performed analytically such that we obtain the identical result as the steady state linear response limit ($F_{ac}=0$) of Eq. (\ref{eq:losstan_eq}), $\tan(\delta)=\frac{4\pi^2 P_{0}p^2}{3\epsilon}$. 
This universal result is a consequence of the Fermi Golden Rule in a linear response limit, which does not depend on the nature of the $\delta$-function broadening, determined by either the decoherence rate $\left(\frac{\hbar}{T_{2}}\right)$ or the energy sweep rate ($\hbar\sqrt{v_{0}}$). 

\begin{figure}[h!]
\centering
\includegraphics[width=7cm]{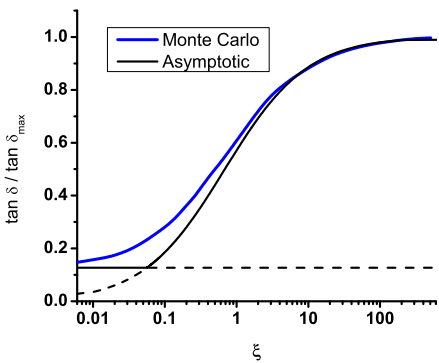}
\caption{The non-equilibrium loss tangent as a function of the dimensionless field sweep rate, $\xi=\frac{2v_{0}}{\pi\Omega_{R0}^{2}}$, in the case 
 $\Omega_{R0}^{2}T_{1,{\rm min}}T_{2,{\rm min}}=81$. The thick blue line shows the Monte-Carlo averaged solution of density matrix equations. The solid black line shows the asymptotic behavior which includes the Landau-Zener solution (Eq. (\ref{eq:losstan_neq})), with the constant loss resulting from steady state behavior. The dashed lines extend Landau-Zener and steady state behaviors to the whole domain of sweep rates.}
\label{fig:Fig3NoneqDelta}
\end{figure}

In opposite (adiabatic) regime, $v_0<<\Omega_{R0}^2$, one can estimate the integral in Eq. (\ref{eq:losstan_neq}) with logarithmic accuracy, meaning that $\int_{0}^{1}dx\frac{1-e^{-ax}}{x}\approx \ln(a)$,  as 
\begin{equation}
\tan(\delta_{ad})=\frac{2\pi^{2} P_{0}p^2}{\epsilon}\frac{2v_{0}}{\pi\Omega_{R0}^{2}}\ln\left(e^{-1/4}\frac{\pi \Omega_{R0}^2}{2v_{0}}\right).   
\label{eq:ad_slowrate}
\end{equation} 
The intermediate regime, $v_{0} \sim \Omega_{R0}^{2}$, can be studied only numerically. 
The result of a numerical calculation of the Landau Zener formula of Eq. (\ref{eq:losstan_neq}) is shown in Fig. \ref{fig:Fig3NoneqDelta} for dielectric losses as a function of the dimensionless sweep rate $\xi=\frac{2v_{0}}{\pi\Omega_{R0}^{2}}$. This result is shown with the corresponding steady state non-linear loss for $\Omega_{R0}T_{1,min}T_{2,min}=81$ by a black line at $0.127$. The two calculations do not agree in the asymptotic slow sweep limit, because the Landau-Zener calculations ignore relaxation and decoherence processes during resonant passage.


Relaxation and decoherence processes must affect dielectric losses at small field sweep rates \cite{LZDecoh}. In fact in the case of fast relaxation $\Omega_{R0}^{2}T_{1,{\rm min}}T_{2,{\rm min}} \ll 1$ one can expect that the linear regime result will be valid at all bias field sweep rates (here the time $T_{2,{\rm min}}$ stands for the inverse maximum decoherence rate including all possible decoherence channels though the results are presented for the low temperature limit $T_{2}\approx 2T_{1}$).
In the opposite, strongly nonlinear limit the microwave absorption should be saturated for small sweep rates, $v_{0}$, and should collapse to the equilibrium case of Eq. (\ref{eq:losstan_eq}). In the steady state equilibrium ($v_{0}=0$) the non-linear microwave absorption comes from the energy domain $\mid E-\hbar\omega\mid \leq \hbar \Omega_{R0} \sqrt{\frac{T_{1,{\rm min}}}{T_{2,{\rm min}}}}$ \cite{Hunklinger,Faoro}. If during the time $T_{1,{\rm min}}$ the change of TLS energy due to bias field sweep, $\delta E \sim \hbar v_{0}T_{1,{\rm min}}$, is small compared to the size of the domain, then one can ignore the field sweep and use the equilibrium result, Eq. (\ref{eq:losstan_eq}). Indeed, at the crossover $v_{0}\sim \frac{\Omega_{R0}}{\sqrt{T_{1,{\rm min}}T_{2}}}$ the equilibrium non-linear loss 
tangent given by Eq. (\ref{eq:losstan_eq}) and the non-equilibrium loss tangent given by Eq. (\ref{eq:ad_slowrate}) become equal to each other within the accuracy of a logarithmic factor, on the order of unity. Thus one can qualitatively approximate the dielectric loss behavior for different Landau-Zener parameters, expressed through the dimensionless field sweep rate  $\xi=\frac{2v_{0}}{\pi\Omega_{R0}^{2}}$ and nonlinearity parameter $\eta\approx \frac{1}{\Omega_{R0}\sqrt{T_{1,{\rm min}}T_{2,{\rm min}}}}$ by either using the steady state limit, Eq. (\ref{eq:losstan_eq}), for small sweep rates, $\xi \ll \eta$, or Landau-Zener relaxation-free limit, Eq. (\ref{eq:losstan_neq}). The latter regime can be characterized by the asymptotic behavior for $\xi\ll 1$ (Eqs. (\ref{eq:ad_slowrate})) or by the large bias rate limit in the opposite case (see Fig. \ref{fig:Fig3NoneqDelta}).


We calculated the non-equilibrium dielectric losses using a full numerical solution to the Bloch equations for each TLS \cite{Hunklinger,ab1,Clare2}. Monte-Carlo integration of the results over the TLS distribution are shown by the thick blue line in Fig. \ref{fig:Fig3NoneqDelta}. The result is consistent with the predicted behavior in the fast sweep rate limit and the steady state results of the slow sweep rate limit.



Using this theory one can experimentally extract the dipole moment $p$, the TLS density $P_0$, and their relaxation time, $T_{1}$, separately, in the low temperature limit under consideration ($T_{2}\approx 2T_{1}$). Experiments can create a known bias sweep, $\frac{dF_{bias}}{dt}$, and ac field $F_{ac}$ and find the TLS dipole moment  $p$ that correctly sets the Landau Zener parameter, $\xi$, to agree with Fig. \ref{fig:Fig3NoneqDelta}.  The dimensionless parameter, $\frac{P_{0}p^2}{\epsilon}$, can be found independently from the loss tangent measurements either in the intrinsic equilibrium regime, or in the strong nonequilibrium limit found for $\xi>>1$, which with the above information of $p$ allows one to separately find $P_0$ and $T_{1}$. Recent experiments with microwave ac and field bias contain the needed experimental regimes $\xi = ~10^{-6} - ~10^{2}$   and $\eta \gg 1$. A full comparison of these experiments to this theory will be completed in a separate work.

	The theory is restricted to the low temperature limit where $k_{B}T \ll \hbar\omega$. 
At higher temperatures and low bias rate $v_{0}T_{1,{\rm min}} \ll k_{B}T$ the result remains applicable after multiplying the losses by the thermal population difference factor $\tanh\left(\frac{\hbar\omega}{2k_{B}T}\right)$. At larger bias rates the population difference cannot equilibrate to the instantaneous energy $E(t)$, and the loss is determined by an earlier (and higher) energy $~E(t-T_1)$. As a result the non-equilibrium loss tangent can exceed its steady state linear response limit if the TLSs are already thermally excited.

It is interesting that in the adiabatic regime many TLS are brought into their excited state creating a remarkable population inversion as in the rapid adiabatic passage regime leading to the phonon enhancement \cite{Sas1,Sas2}. 

In conclusion we propose a theory to explain the effect of a sweeping electric field bias on the ac resonant loss in an amorphous dielectric. 
If the field sweep rate is very fast the loss tangent reaches a universal value even in the strongly non-linear regime of 
high amplitude ac fields, in agreement with recent experimental observations \cite{K2}. At slower bias sweep rates a  strongly non-linear regime takes place, in which the loss tangent decreases with decreasing the sweep rate until the saturation at the steady state non-linear limit Eq. (\ref{eq:losstan_eq}). 
This nonequilibrium loss tangent theory can be used to interpret and extract many TLS properties in various contexts and provide a better physical insight into them.

This work is supported  by the Louisiana Board of Regents Pfund project no. 
PFUND-330. Authors acknowledge Christian Enss, Moshe Schechter, Andreas Fleischmann	 and Sergiy Gladchenko for useful discussion.

\end{document}